# ANALYSIS & PREDICTION OF SALES DATA IN SAP-ERP SYSTEM USING CLUSTERING ALGORITHMS

S.HanumanthSastry and Prof.M.S.PrasadaBabu

Department of Computer Science & Systems Engineering, Andhra University, Visakhapatnam, India

*ABSTRACT*

*Clustering is an important data mining technique where we will be interested in maximizing intracluster distance and also minimizing intercluster distance. We have utilized clustering techniques for detecting deviation in product sales and also to identify and compare sales over a particular period of time. Clustering is suited to group items that seem to fall naturally together, when there is no specified class for any new item. We have utilizedannual sales data of a steel major to analyze Sales Volume & Value with respect to dependent attributes like products, customers and quantities sold. The demand for steel products is cyclical and depends on many factors like customer profile, price,Discounts and tax issues. In this paper, we have analyzed sales data with clustering algorithms like K-Means&EMwhichrevealed many interesting patternsuseful for improving sales revenue and achieving higher sales volume. Our study confirms that partition methods like K-Means & EM algorithms are better suited to analyze our sales data in comparison to Density based methods like DBSCAN & OPTICS or Hierarchical methods like COBWEB.*

## 1. INTRODUCTION

Sales and distribution function of an enterprise handles many important business processes like preparing annual/monthly marketing plans, developing pricing strategies, handling service contracts, manage logistics for material dispatches and maintaining customer relationships [1].Steel plant produces more than 3300 products out of which 1500 are main products and the remaining are miscellaneous products having different length variations. Pre-sales information is used to plan and evaluate marketing & sales strategies and as a basis for establishing long term relationship with customers. For our study, we have sourced information relating to customer demands for different steel products measured in terms of both Value & Volume. Sale order is an important document in ERP systems that captures customers request for goods [2]. Sale orders are categorized into orders such as special steels, domestic, export, stockyard, by products, region wise, LTC (Long term contract), Project Sales etc. Customersfor the products could belong to Government/Private/Public sector and also Corporations like Railways, Airports, Defense, Infrastructure, and Rural/Urban development etc. Customer floats enquires on products, quantity, quality, dimensions, delivery dates, destination etc [3]. Based on the customer demand, new production orders are generated and sometimes alternate order sizes are offered, if the original customer order is not feasible for the specified size, grade.Sales targets and market sentiments are also considered for fixing the price. In this paper, we have applied clustering techniques to divide the sales data of enterprise into clusters that are meaningful, useful or both. The rest of this paper is organized as follows- Section 2 describes sales analysis dependent issues, Section 3 describes various clustering techniques & algorithmic framework, Section 4 describes Implementation steps, Section 5 presents Results & Discussion andSection 6 provides conclusions on the paper.

## 2. SALES DATA ANALYSIS ISSUES

Marketing practices are driven by technologies that enable closer engagement with customers and also the acquisition and analysis of more and more data on customers [2].Planning and Execution of Marketing activities are based on pre-defined policies on various types of sales. Every year, based on the existing market conditions all the policies are reviewed and refined to suit the existing conditions. These policies guide the planning process in marketing function. To establish loyal customer base, the Long Term Contracts (LTC) policy is adopted at the discretion of the management. In order to promote the sales, Credits are given to trustworthy customers in various segments.For maintaining consistent sales irrespective of demand, company enters in to an agreement with some of the regular customers with committed quantities in the beginning of the year [3]. The criteria of allocating the tonnage for each customer may vary based on the policy afore mentioned and can every year depending on the market demand. Also credits are extended to customers as per credit policy. The other crucial aspects of sales function are discussed here.

**2.1 Product Attributes**: In order to identify the products Mill wise, Product group wise, size wise, grade wise and length wise, a 15 digit code is prepared with identified digits for each of the above characteristics. There is a predetermined pattern for giving codes for various mill products, various sizes, grades and lengths. If a new product is manufactured with different size or grade, the product code is generated as follows:

```
X   XX    XXX    XXXXX   XXXX
(1) (2)   (3)    (5)     (4)    where
```
1. Mill (WRM, LMMM etc)
2. Product Group (Rebars, Structurals, etc)
3. Sizes (Diameter, section of angles/channel etc)
4. Grade or Quality
5. Length

**2.2 Annual Sales Targets** (product category wise, for example Pig iron and steel) are based on periodic assessment, availability of material and market demand andbroken down into bi-monthly, Quarterly targets. Traders and actual users take the material and take up the final selling activities [2].

**2.3 Price Fixation**: Pricing is typically based on *conditions*. Here we define conditions as a set of circumstances that are applied when we calculate the price. To cite an example, a particular customer can order a specified quantity of product on a particular day. The variable factors here are the customer, the product, the order quantity, and the date which together determines the final price the customer gets. Master data pertaining to this is stored is stored as*condition records* in database. The pricing also considers seasonal effects on the demand like monsoon, festival seasons, year beginning and year ending and export plans of company as well as competitors and import plans of major customers, changes in duties and railway freight etc. After fixation of prices, the enquiries are sent to all existing customers indicating the price material wise and grade wise by mail. After getting confirmation from customers, the clearance for availability of material is taken from Planning and dispatch section and thereafter Contracts are issued.

**2.4 Net Sales Realization(NSR)** values should be improved for each quarter as per set targets. Often it is observed that Delivery Order (DO) is issued for more quantity than stock available, in anticipation of higher production, and because of change in priorities there could be lesser production.Achieving product mix adherence i.e. % adherence to target product mix as defined in revenue budget, is a priority. Price of some of the products where chemical composition plays a role is decided just before delivery.

**2.5 Handling quality complaints**:There would be refunds against defective materials received by customers. Lead time for settlement of complaint, which is the average time taken from date of complaint to the date on which the complaint is fully closed, need to be reduced.

**2.6 Launch of New Products**: New product development is an opportunity for growth and sustenance of any organization. Efforts are made to develop new products which can penetrate in to new markets and contribute to Higher Net Sales Realization (NSR).Based on quality feedback from the market segment, new and niche markets are developed. Acceptability of new products developed is measured in terms of % sales of new products developed on total sales.Figure 1 summarizes above dependent issues on sales function [4].

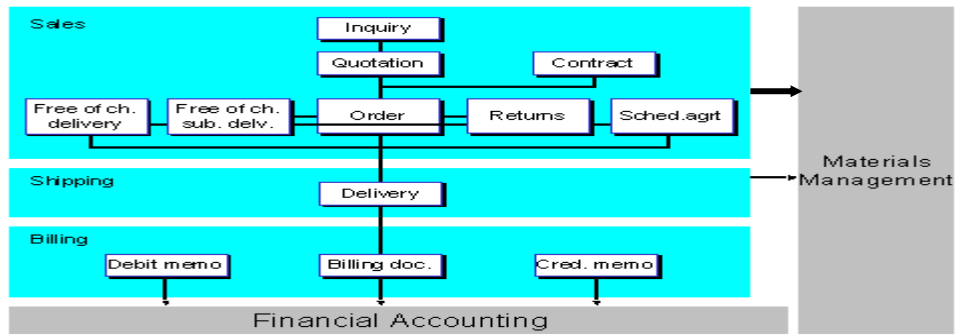

Figure 1: Sales Organization Interfacewith Shipping & Billing Functions

Top executives of industry are challenged to move the company perspective from a product-oriented to service-oriented organization [3]. From our study it is observed that number of product grades developed over the years is more than 90, but grades in demand are not more than 20-25. To understand these issues from customer perspective, we have utilized learning techniques of data mining as discussed in next section.

## 3. CLUSTER ALGORITHMS

Clustering Algorithms available in WEKA with their respective functions are listed in Table 1.The input to cluster algorithms is a set of instances. These instances are the things that are to be International Journal of Computational Science and Information Technology (IJCSITY) Vol.1, No.4, November 2013classified or associated or clustered. We can characterize instances by the values of a specified set of predefined attributes [5].

Table 1: Clustering Algorithms in Weka

| Name of Algorithm | Function |
|---|---|
| CLOPE | Fast Clustering of Transactional data |
| COBWEB | Implements Cobweb &Classit Clustering Algorithms |
| DBScan | Nearest-neighbor-based clustering that automatically determines no of clusters |
| EM | Cluster using Expectation Maximization |
| FarthestFirst | Uses Farthest First Traversal Algorithm |
| FilteredClusterer | Runs a clusterer on filtered data |
| HierarchicalClusterer | Agglomerative hierarchical clustering |
| MakeDensityBasedCluster | Wrap a cluster to make it return distribution and density |
| Optics | Extension of DBScan to hierarchical clustering |
| sIB | Clustering using the sequential information bottleneck algorithm |
| SimpleKMeans | Cluster using the k-means method |
| XMeans | Extension of k-means |

### 3.1 Hierarchical clustering methods

In Hierarchical clustering, we group data items into a tree consisting of a definite number of clusters. Hierarchical clustering methods can be further classified as either *agglomerative* or *divisive*, depending on whether the hierarchical decomposition is formed in a bottom-up (merging) or top-down (splitting) fashion [6]. Hierarchical methods examines all the clusters present so far at each stage of merging, the clustering methods we examined work incrementally, instance by instance. At any stage the clustering forms a tree with instances at the leaves and a root node that represents the entire dataset. At the beginning of clustering, the tree consists of only the root node. Instances are added one by one, and the tree is updated appropriately at each stage. The key to deciding how and where to update the tree is a quantity called *category utility* that measures the overall quality of a partition of instances into clusters. The important algorithms in hierarchical clustering are -

**3.1.1 Agglomerative Clustering** (**Bottom-up**): Cobweb (Nominal Attributes) &Classit (Numeric Attributes). Cobweb &Classit rely on hierarchical group of data instances. They deploy a unique measure of cluster quality known as *category utility*. It measures distance between any 2 clusters and treats any instance as a cluster in its own right, then finds the two closest clusters, merges them, and keep on doing this until only one cluster is left. The record of merging forms a hierarchical clustering structure—a binary dendrogram. These methods are sub-divided into – Agglomerative and Divisive hierarchical clustering. In Agglomerative clustering each object represents a cluster of its own [7].

**3.1.2 Cobweb algorithm** – it always compares the best host, adding a new leaf, merging the two best hosts, and splitting the best host when considering where to place a new instance. Parameters for classes implementing Cobweb &Classit are acuity & cutoff.*HierarchicalClusterer*implements agglomerative generation of hierarchical clusters [9].

### 3.2 Partition Clustering Methods

In Partition clustering clusters are created for optimizing a *predetermined criterion*. The criteria adopted for our study is dissimilarity function which is essentially based on object distances

98

within and across clusters. The grouping is done as follows – objects within a same cluster are grouped as "similar" and objects belonging to different clusters are grouped as "dissimilar". The object attributes play a crucial role in this grouping of clusters. Given *D*, a data set of *n* objects, and *k*, the number of clusters to form, a partitioning algorithm organizes the objects into *k* partitions ($k \leq n$), where each partition represents a cluster [6]. Important partition clustering algorithms are –

**3.2.1 K-Means Algorithms**: When considering splitting a cluster, there is no need to consider the whole tree, just look at those parts of it that are needed to cover the cluster. *XMeans* implements an extended version of *k*-means by Moore and Pelleg(2000). It uses a Bayesian information criterion for selecting the number of clusters, and can use *k*D-trees for speed .We can specify the distance function to use, the minimum and maximumnumber of clusters to consider, and the maximum number of iterations to perform. Farthest First is modeled on k-means & implements the farthest-first traversal algorithm. *MakeDensityBasedClusterer* is a metaclusterer that wraps a clustering algorithm to make it return a probability distribution and density. To each cluster and attribute it fits a discrete distribution or a symmetric normal distribution [7].

**3.2.2 EM Algorithm**: Instances pertaining to input data sets are assigned to clusters based on criteria called *expectation*. This is similar to finding out the class of an instance which is previously unknown. Estimating the input parameters from the already classified instances is referred as *maximization.* This is similar to finding out the probabilities of attribute values from the training data instances. In EM algorithm we assign classes based on *probabilistic* calculation. This is in contrast to algorithms where instances are assigned *categorically*. The EM procedure guarantees finding model parameters that have equal or greater likelihood at each iteration. The key question to be answered here is whether the higher probability of parameter estimates will help in increasing the cluster classification accuracy [5] [6].

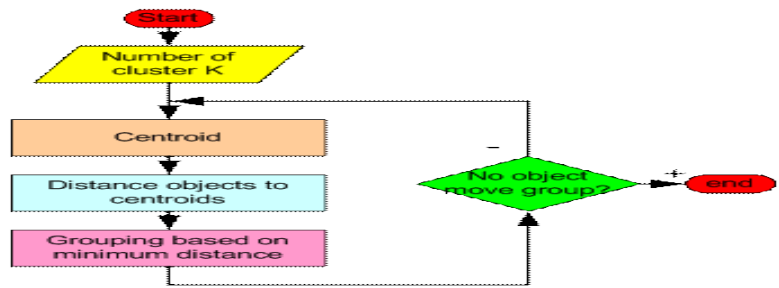

Figure 2: Illustration of K-Means Algorithm

Step 1. Begin with a decision on the value of k = number of clusters

Step 2. Put any initial partition that classifies the data into k clusters. You may assign the training samples randomly, or systematically as the following - Take the first k training sample as single-element clusters & assign each of the remaining (N-k) training samples to the cluster with the nearest centroid. After each assignment, the centroid of the gaining cluster is recomputed.

Step 3.Take each sample in sequence and compute its distance from the centroid of each of the clusters. If a sample is not currently in the cluster with the closest centroid, switch this sample to that cluster and update the centroid of the cluster gaining the new sample and the cluster losing the sample.

Step 4. Repeat step 3 until convergence is achieved, that is until a pass through the training sample causes no new assignments.

Since we are not sure about the location of the centroid, we need to adjust the centroid location based on the current updated data. Then we assign all the data to this new centroid. This process is repeated until no data is moving to another cluster anymore. Mathematically this loop can be proved to be convergent. The convergence will always occur if the following condition satisfied:

1. Each switch in step 2 the sum of distances from each training sample to that training sample's group centroid is decreased.

2. There are only finitely many partitions of the training examples into k clusters.

If a clustering method were used to label the instances of the training set with cluster numbers, that labeled set could then be used to train a rule or decision tree learner [7]. The resulting rules or tree would form an explicit description of the classes.

### 3.3 Density Based Methods

This discovers clusters with arbitrary shapes & typically regard clusters as dense regions of objects in the data space that are separated by regions of low density (representing noise). DBSCAN grows clusters according to a density-based connectivity analysis. OPTICS extends DBSCAN to produce a *cluster ordering* obtained from a wide range of parameter settings. DENCLUE clusters objects based on a set of density distribution functions[7][8].

**3.3.1 DBScan** uses the Euclidean distance metric to determine which instances belong together in a cluster (Ester et al., 1996), but, unlike *k*-means, it can determine the number of clusters automatically, find arbitrarily shaped clusters, and incorporate a notion of outlier. Here we specify cluster as having a minimum number of points. For every pair of points, either can lie within a user specified distance ($\varepsilon$) of each other point which are then inter connected in the specific cluster. Each of these points lies within a distance $\varepsilon$ of the next point in the cluster. Smaller values of $\varepsilon$ yield denser clusters because instances must be closer to one another to belong to the same cluster. Depending on the value of $\varepsilon$ and the minimum cluster size, it is possible that some instances will not belong to any cluster. These are considered outliers [5] [9] [10].

**3.3.2 OPTICS** is an extension of *DBScan* to hierarchical clustering (Ankerst et al., 1999). It imposes an ordering on the instances, which, along with two-dimensional visualization, exposes the hierarchical structure of the clusters. The ordering process places instances that are closest to one another, according to the distance metric, beside one another in the list. It annotates each adjacent pair of instances with the "reachability distance", which is the minimum value of $\varepsilon$ that allows the pair to belong to the same cluster. The clusters become apparent when ordering is plotted against reachability distance. Because instances in a cluster have low reachability distance to their nearest neighbors, the clusters appear as valleys in the visualization. The deeper the valley, the denser the cluster [6] [9].

**3.4** *sIB* is an algorithm designed for document clustering that uses an information theoretic distance metric (Slonim et al., 2002). The number of clusters to find and the maximum number of iterations to perform can be specified by the user [5].

The implementation steps forclustering algorithms in Weka are discussed in next section.

## 4   IMPLEMENTATION STEPS

Weka clustering algorithms were run on Windows 2008 server with Intel dual core 1.83 GHz processor & 4 GB RAM with physical address extension. Classifiers are used to improve classification [11] of unlabeled data. Sales data analysis attributes and detailed steps of implementation are listed below.

Table 2: Attributes Modeled in Clustering algorithms

| Attribute | Attribute Description (Quantity in Tons; Sales Value in INR) |
|---|---|
| Product_CD | Product Code |
| Prod_Desc | Product Description |
| Customer_CD | Customer Code |
| No_of_Records | No of times a single product bought by each customer in the year |
| Quantity_sold | Sales Volume for a single product against each customer in the year |
| Sales_value | Sales Value for a single product against each customer in the year |

1. **Importing SQL query results on aggregated data of relevant source fields**: As source data is at a higher level of granularity, we have formulated a SQL query at database level to extract annual aggregated data from relevant fields like Products, Customers and used the query results in Weka Explorer.

2. **Choosing the Classifier**: We have used ZeroR classifiers to learn classes from a small labeled dataset and then extended it to a large unlabeled datasets for different algorithms like EM iterative clustering algorithm.

3. **Evaluation of attributes**: Chosen Evaluator – CfsSubsetEval ; Search Method: BestFirst in full training mode

4. **Running Clustering algorithms**: We have used 66% of data for training and the remaining for clustering on following clustering algorithms & also run 'Classes to cluster evaluation'
5. **Visualization of results:** Graphical display of algorithm results in 2D scatters plot diagrams[17].

## 5   RESULTS & DISCUSSION

The results of ZeroR Classifier on 10 Fold Cross-validation for numeric attributes Sales Volume, Sales Value & No of Records is given in table 3.Attribute evaluator *CfsSubsetEval* is indicative of

International Journal of Computational Science and Information Technology (IJCSITY) Vol.1, No.4, November 2013

the worth of a subset of attributes. This evaluator considers the individual predictive ability of each feature. Also the degree of redundancy between the features is accounted [12] [13]. For our study, we have preferred subsets of features that are highly *correlated* within a class. Classes with low *inter-correlation* values are also preferred [14]. For a total number of 17 subsets evaluated, the values for '*MERIT OF BEST SUBSET FOUND*' = 0.527 &*SELECTED ATTRIBUTES* – 4:1 (No_of_Records). For all 18901 instances, Simple k-Means algorithm iterated 17 No of times & has shown '*WITHIN CLUSTER SUM OF SQUARED ERRORS*' as 52873.298. Cluster#0 & Cluster#1 has 83% & 17% instance values respectively as shown in table 4.

Table 3: Numeric Attributes of Sales Data classified with ZeroR Classifier

| Selected Attribute | Sales Volume | Sales Value | No of Records |
| --- | --- | --- | --- |
| ZeroR Predicted Class Value | 2489929.417 | 116.597 | 3.194 |
| Correlation Coefficient | -0.0247 | -0.0262 | -0.0109 |
| Mean Absolute Error | 2884054.091 | 160.337 | 2.490 |
| Root Mean Squared Error | 8460917.861 | 3047.589 | 4.808 |
| Relative Absolute Error | 100% | 100% | 100% |
| Root Relative squared error | 100% | 100% | 100% |
| Total No. of Instances | 18901 | 18901 | 18901 |

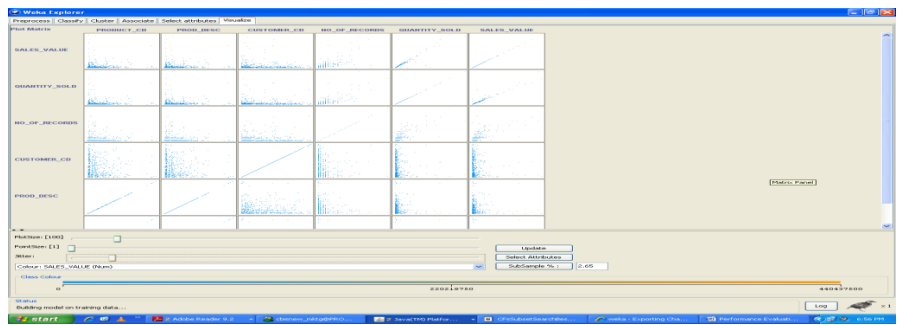

Figure 3: Visualization of Attribute Matrix Panel in Weka Explorer

Table 4: Cluster Centroids for Simple K-Means algorithm

| Attribute | Full Data (18901) | Cluster#0 (15655 Instances) | Cluster#1 (3246 Instances) |
| --- | --- | --- | --- |
| PRODUCT_CD | 603008041210000 | 503016041210000 | 603012041210000 |
| PROD_DESC | TMT Rebar Coil 8 IS1786 Fe 500 | TMT Rebar 16 IS1786 Fe 500 | TMT Rebar Coil 12 IS1786 Fe 500 |
| CUSTOMER_CD | 120D200004 | 000A140102 | 120D200004 |
| NO_OF_RECORDS | 3.1944 | 1.991 | 8.9982 |
| QUANTITY_SOLD | 116.5965 | 32.2141 | 523.561 |
| SALES_VALUE | 2489929.4165 | 1131837.9764 | 9039812.8099 |

More outlier transactions are grouped in Cluster 1 as seen in figure 1 below. Also '***Classes to Cluster Evaluation***' run has shown better accuracy with *products* rather than *customers* attribute.

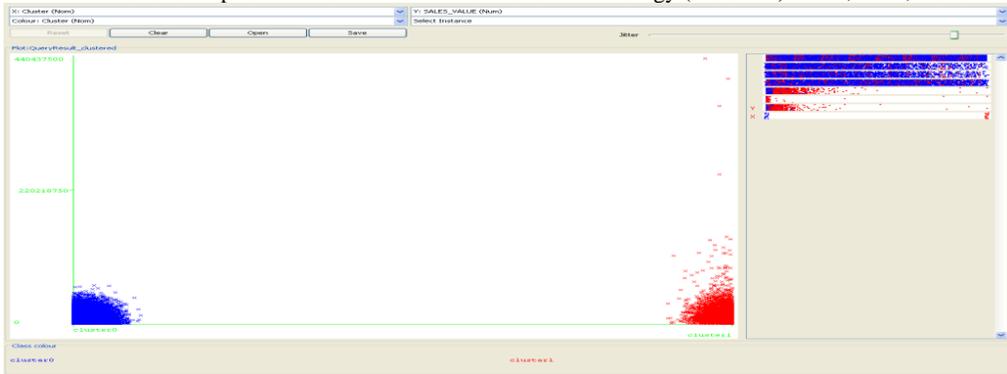

Figure 4: Simple K-Means Algorithm: Clusters vs. Sales Value

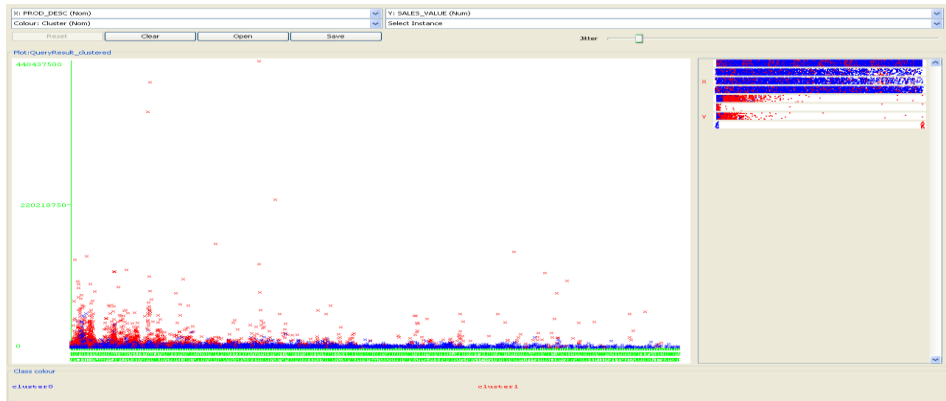

Figure 5: Simple K-Means Algorithm: Products vs. Sales Value

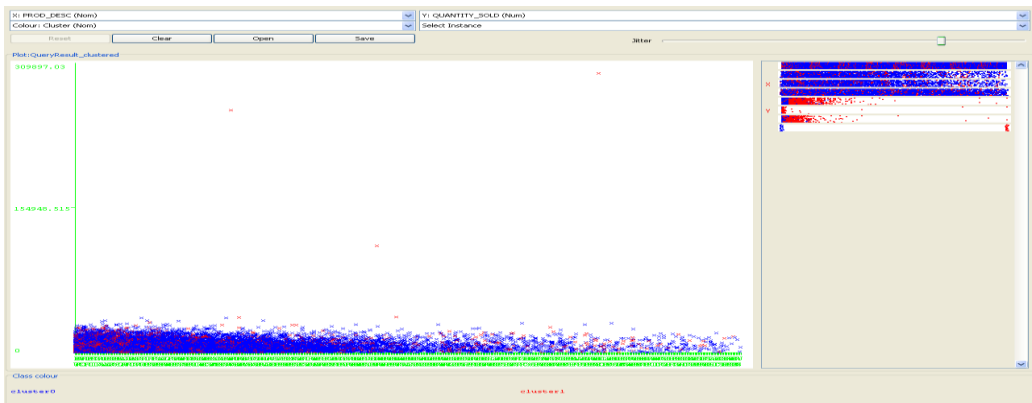

Figure 6: Simple K-Means Algorithm: Products vs. Sales Volume

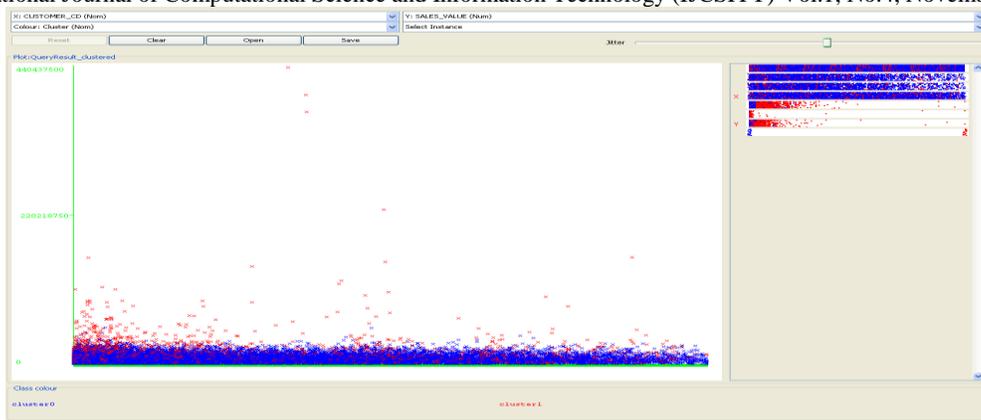

Figure 7: Simple K-Means Algorithm: Customers vs. Sales Value

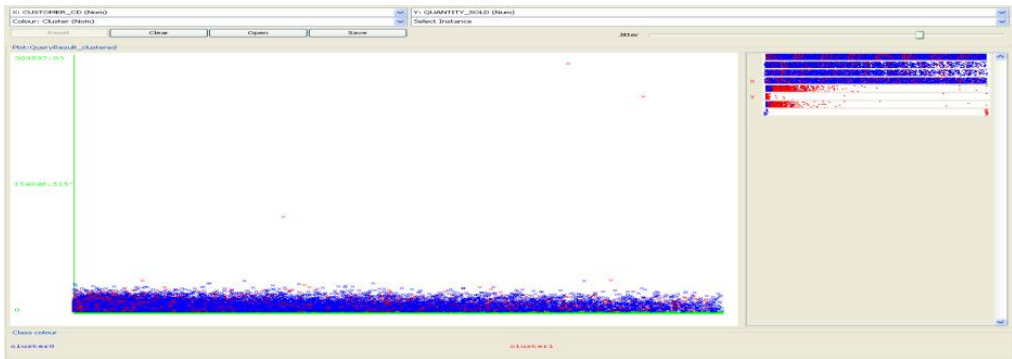

Figure 8: Simple K-Means Algorithm: Customers vs. Sales Volume

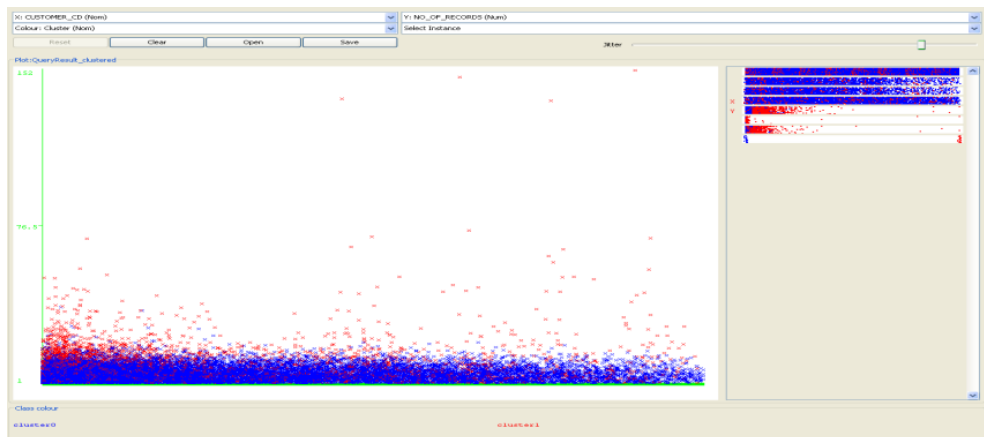

Figure 9: Simple K-Means Algorithm: Customers vs. No_of_Records

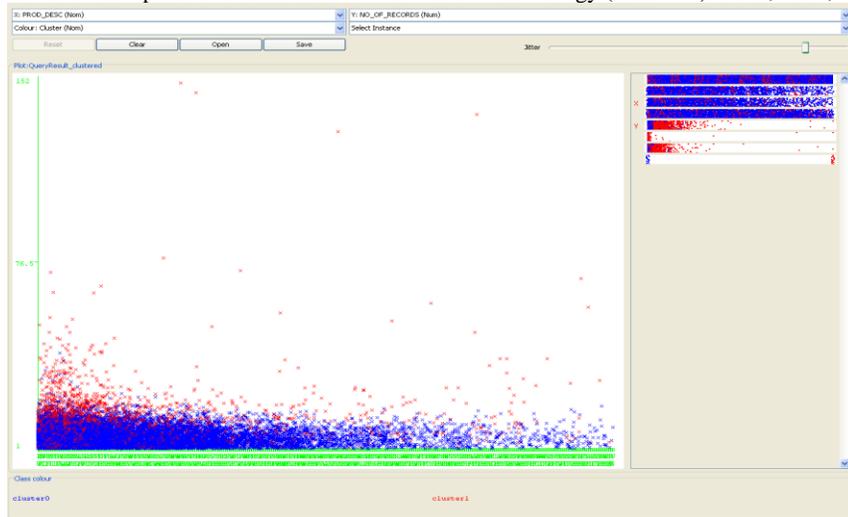

Figure 10: Simple K-Means Algorithm: Products vs. No_of_Records

**DBScanalgorithm** with parameters ε = 0.9 &minPoints = 6, has not generated any clusters and treated all instances as 'NOISE'. Hence DBSCan algorithm is not found suitable for analyzing input sales data.

**OPTICS algorithm** with parameters ε = 0.9 &minPoints = 6, has not generated any clusters and treated all instances as 'UNDEFINED' [16]. Hence OPTICS algorithm is not suited for the type of sales data we have used. COBWEB algorithm ran for too long with parameters acuity = 1.0, cutoff = 0.00282 & seed = 42. We have not considered these results either.

**EM algorithm** has created 4 clusters on cross-validation with values for prior possibilities as: Cluster0-0.39, Cluster1-0.04, Cluster2-0.32 and Cluster3-0.24.Total No of Instances in clusters 0,1,2,3 are 10157.0809,3512.5079, 8887.0809 and 7340.3303 respectively. It has taken 1431.83 seconds to build model on full training data. Log-likelihood is the main evaluation measure for comparison purposes [15] & has result value for model: -38.04232. The number of instances assigned to each cluster[16] & their respective Percentages when the learned model is applied to the data as a classifier is shown in table 5.

Table 5: EM Algorithm Clustering Results

| **Numeric Attribute** | | **Cluster0** | **Cluster1** | **Cluster2** | **Cluster3** |
|---|---|---|---|---|---|
| No_of_Records | Mean | 1.143 | 17.95 | 2.29 | 5.25 |
| | Std.Dev | 0.35 | 14.32 | 1.19 | 3.26 |
| Quantity_Sold | Mean | 6.09 | 1929.19 | 26.99 | 113.27 |
| | Std.Dev | 3.65 | 15045.98 | 12.07 | 76.72 |
| Sales_value | Mean | 225617.82 | 27113906.80 | 986642.45 | 4058298.08 |
| | Std.Dev | 135610.07 | 32195752.77 | 438024.16 | 2722933.25 |
| Clustered Instances | | 7451 | 765 | 6142 | 4543 |
| % of Clustered Instances | | 39% | 4% | 32% | 24% |

International Journal of Computational Science and Information Technology (IJCSITY) Vol.1, No.4, November 2013

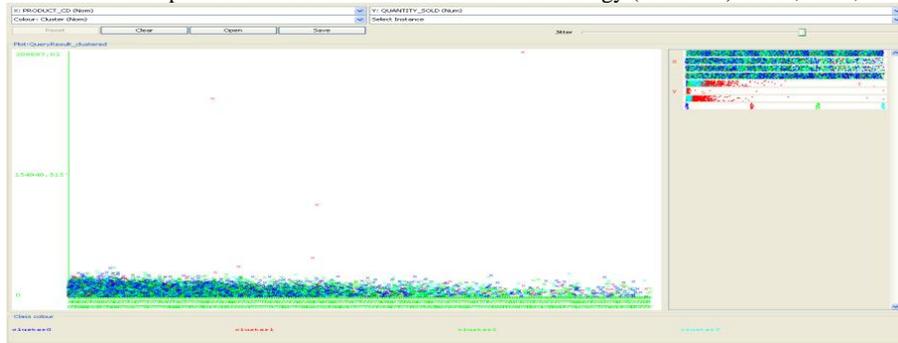

Figure 11: EM Algorithm: Products vs. Sales Volume

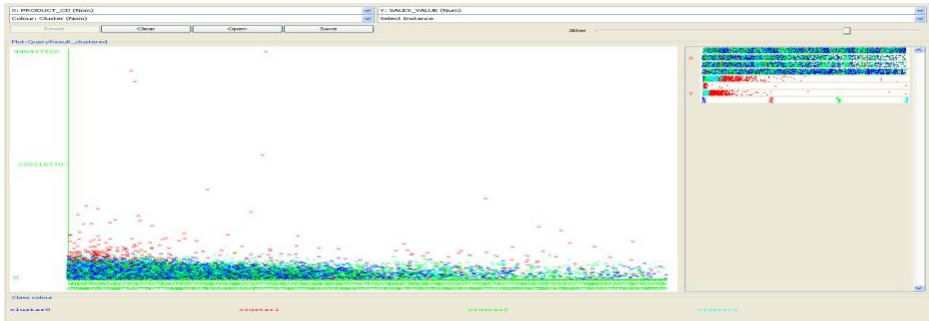

Figure 12: EM Algorithm: Products vs. Sales Value

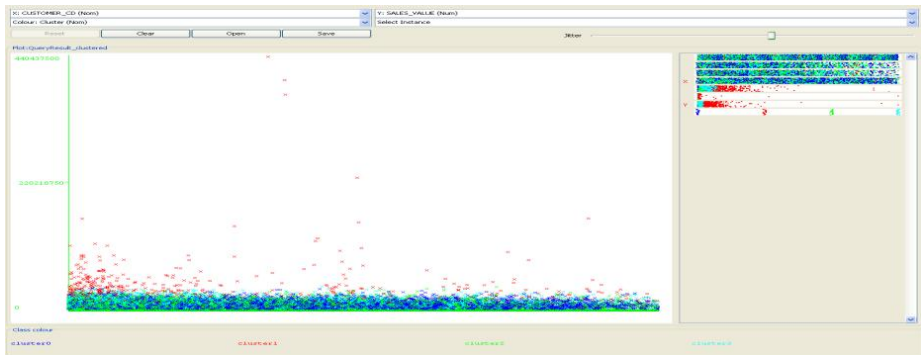

Figure 13: EM Algorithm: Customers vs. Sales Value

International Journal of Computational Science and Information Technology (IJCSITY) Vol.1, No.4, November 2013

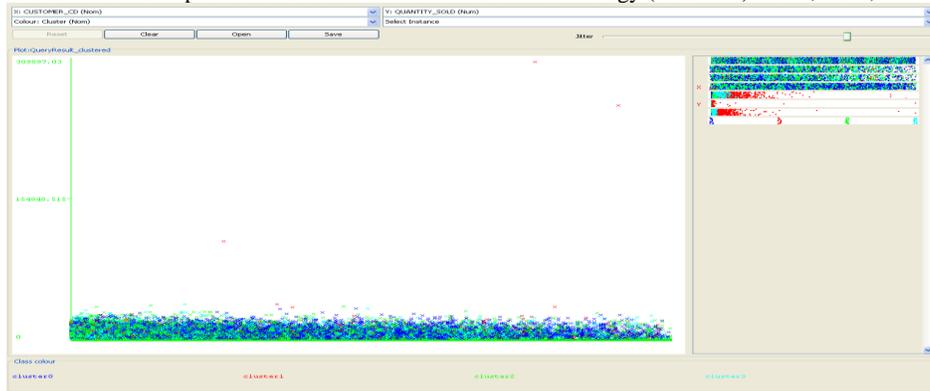

Figure 14: EM Algorithm: Customers vs. Sales Volume

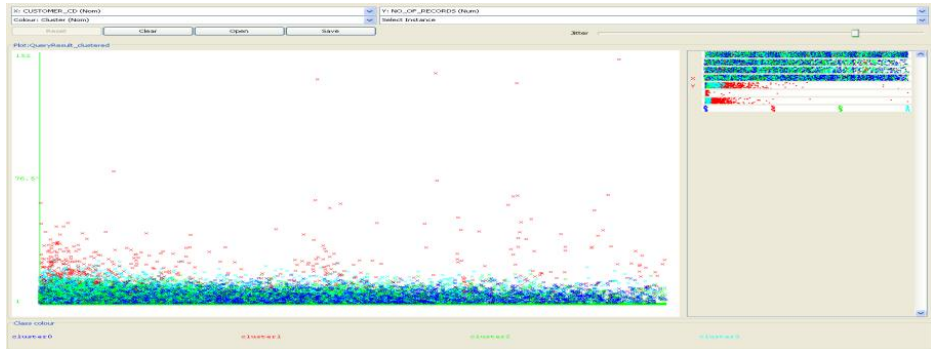

Figure 15: EM Algorithm: Customers vs. No_of_Records

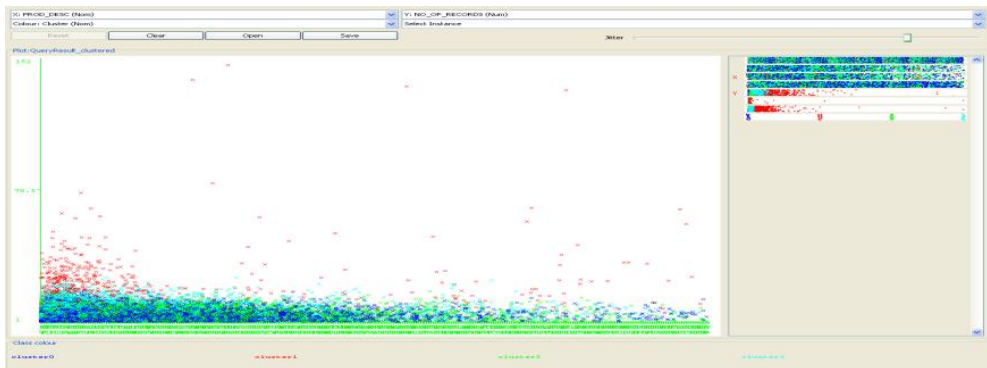

Figure 16: EM Algorithm: Products vs. No_of_Records

Table 6:Results of FarthestFirst Algorithm (2 Clusters - 100% of instances are in cluster0)

| Attribute | Cluster0 (18899) | Cluster1 (2) |
|---|---|---|
| PRODUCT_CD | 601010710100000 | 055000800000011 |
| PROD_DESC | WR Coil 10 SAE1010 | BFG SLAG (EXPORT) |
| CUSTOMER_CD | 510A080008 | 010A160162 |
| NO_OF_RECORDS | 6.0 | 134.0 |
| QUANTITY_SOLD | 47.54 | 309897.03 |
| SALES_VALUE | 1688200.28 | 1.146619011E8 |

## 6 CONCLUSIONS

Cluster analysishelps in Market Intelligence for correctly assessing the demand and the net sales realization for the product to be developed. It is also possible to identify lost order opportunities and improve sales volumes for new products. From our analysis it is seen that product 'B.F.G. SLAG' has high export potential. From result graphs, it can be observed that for certain products niche market exists and sales value will also be high. With available sales data we can establish relations between the presale activities like tenders, quotations, enquiries etc with the sale order and post sale activities like pricing, destinations, invoicing, refunds, returns etc. This helps in analyzing monthly planned vs. actual sales and also an ability to analyze product wise, customer wise sales over a year and further the customer base. The sales data analysis can also be used to achieve higher stock to sales ratio byexploring the possibility of customer wise product wise pricing. The success criterion for any data mining application is accounted in terms of how accurate and acceptable is the implemented solution. This aspect is measured in terms of learned descriptions like new rules framed or inferences drawn which can be utilized by a human user.In this context above analytical results produced by clustering techniques are quite reliable.

**Authors**

**S.HanumanthSastry**Senior Manager (ERP) has implemented SAP-BI Solutions for Steel Industry. He holds M.Tech (Computer Science) from NIELIT, New Delhi and MBA (Operations Management) from IGNOU, New Delhi. His research interests include ERP systems, Data Mining, Business Intelligence and Corporate Performance Management. He is pursuing PhD (Computer Science) from Andhra University, Visakhapatnam (INDIA).

**Prof. M.S. Prasad Babu** obtained his Ph.D. degree from Andhra University in 1986. He was the Head of the Department of the Department of Computer Science & Systems Engineering, Andhra University from 2006-09. Presently he is the Chairman, Board of Studies of Computer Science & Systems Engineering. He received the ISCA Young Scientist Award at the73rd Indian Science Congress in 1986.